\documentclass[pra,twocolumn,letterpaper,amsfonts]{revtex4}

\usepackage{graphicx}

\usepackage{amsthm}
\usepackage{amsmath}
\usepackage{amsfonts}

\usepackage{amssymb}

\usepackage{bm}

\begin{document}

\bibliographystyle{apsrev}

\newcommand{\bra}[1]{\ensuremath{\left \langle #1 \right |}}
\newcommand{\ket}[1]{\ensuremath{\left | #1 \right \rangle}}
\newcommand{\braket}[2]{\ensuremath{\left \langle #1 \right | \left. #2 \right \rangle}}
\newcommand{\tr}{\ensuremath{\mbox{Tr}}}
\renewcommand{\H}{\ensuremath{\mathcal{H}}}
\newcommand{\var}{\ensuremath{ \mbox{var}}}
\newcommand{\hvar}[1]{\ensuremath{ \hat{\mbox{var}} \left ( #1 \right )}}

\title{The Extent of  Multi-particle Quantum Non-locality}
\author{Nick S. Jones}
\affiliation{Dept. of Mathematics, University of Bristol,
University Walk, Bristol, BS8 1TW, UK}
\author{Noah Linden}
\affiliation{Dept. of Mathematics, University of Bristol,
University Walk, Bristol, BS8 1TW, UK}
\author{Serge Massar}
\affiliation{Laboratoire d'Information Quantique and Centre for
Quantum Information and Communication, {C.P.} 165/59, Av. F. D.
Roosevelt 50, B-1050 Bruxelles, Belgium}

\date{\today}

\begin{abstract}
It is well known that entangled quantum states can be nonlocal:
the correlations between local measurements carried out on these
states cannot always be reproduced by local hidden variable
models. Svetlichny, followed by others, showed that multipartite
quantum states are even more nonlocal than bipartite ones in the
sense that nonlocal classical models with (super-luminal)
communication between some of the parties cannot reproduce the
quantum correlations. Here we study in detail the kinds of
nonlocality present in quantum states. More precisely we enquire
what kinds of classical communication patterns cannot reproduce
quantum correlations. By studying the extremal points of the space
of all multiparty probability distributions, in which all parties
can make one of a pair of measurements each with two possible
outcomes, we find a necessary condition for classical nonlocal
models to reproduce the statistics of all quantum states. This
condition extends and generalises work of Svetlichny and others in
which it was shown that a particular class of classical nonlocal
models, the ``separable'' models, cannot reproduce the statistics
of all multiparticle quantum states. Our condition shows that the
nonlocality present in some entangled multiparticle quantum states
is much stronger than previously thought. We also study the
sufficiency of our condition.
\end{abstract}

\maketitle

\section{Introduction}
A natural way of characterizing the correlations present in
entangled quantum states is to attempt to replicate their
measurement statistics using classical models. Bell \cite{B}
showed that classical models respecting relativistic causality -
often called local hidden variable ({\em lhv}) models - cannot
always reproduce quantum statistics. If the classical model allows
unlimited (superluminal) communication between the parties then
the quantum correlations can be reproduced trivially. More refined
studies attempt to understand exactly what superluminal classical
communication will reproduce the quantum correlations. In the
bipartite case, efforts have concentrated on the number of bits of
communication required to reproduce the quantum correlations
\cite{Brass,Stein,Mass,Toner}. In the three party setting
Svetlichny \cite{Svet} showed that, even allowing superluminal
communication between arbitrary pairs of parties cannot reproduce
the results of measurements performed on quantum states. Two
papers independently  generalised Svetlichny's result to an
arbitrary number of parties \cite{Seev,Col}.

In the present work we refine this analysis  by showing that the
class of classical correlations that cannot reproduce the quantum
correlations is much larger than the separable class considered in
 \cite{Svet, Seev, Col}. We show that, not only is the non-locality of
multi-particle quantum correlations stronger than previously
thought, but also there is a way of categorizing non-local
correlations using graphs of communication.

Let us first recall Bell's idea. Consider $m$ parties which each
receive as input a  measurement setting $x_i$ and produce an
output $a_i$. The probability of a certain outcome for a given set
of settings is:\begin{small}
\begin{equation}
P(\vec a| \vec x)=P(a_1,...,a_m|x_1,...,x_m). \label{first}
\end{equation}\end{small}One way to generate such correlations is for the parties to share
an entangled quantum state. Depending on his input, each party
then carries out a measurement on the quantum state. The result of
the measurement is his output. We denote the quantum mechanical
correlations obtained in this way by $P_{qm}$.

The most general way of generating correlations using only
classical resources, without any signalling taking place between
the parties, is for the parties to share a prior random variable
$\lambda$ (often called the hidden variable). Each party then
chooses its outcome depending on its input and on $\lambda$. The
set of correlations produced using such local hidden variable
models has the form:\begin{small}
\begin{eqnarray}
P_{lhv}(\vec a|\vec x)=\begin{tiny}\int
\end{tiny} d\lambda
P(a_1|x_1,\lambda)P(a_2|x_2,\lambda)..P(a_m|x_m,\lambda)\rho(\lambda),
\label{Bell}
\end{eqnarray}\end{small}where $\rho(\lambda)$ is a probability distribution over the
variables $\lambda$.

Note that neither the quantum nor the {\em lhv} models allow
signalling, since in neither of the models does any communication
take place between the parties after they receive their input.
Bell's central result was to show that some quantum correlations
cannot be reproduced by {\em lhv} models. This is proved by
introducing an inequality -called a ``Bell inequality''- which
must be satisfied by the {\em lhv} models but which is violated by
the quantum correlations. See for instance \cite{Wolf,Zuk} for all
``correlation inequalities'' in the multipartite case.

Svetlichny \cite{Svet} generalised and refined Bell's result in
the three party setting by showing that some three-party quantum
correlations cannot be reproduced classically, even if
communication, about settings and results, is allowed between a
pair of parties. The two parties in communication need not be
fixed in advance, but can be chosen with probabilities $p_i$. The
correlations considered by Svetlichny are thus of the form:
\begin{eqnarray}
&&P_{Svet}(\vec a|\vec x)=\nonumber\\&&\int d\lambda\big[
p_1\rho_1(\lambda)P_1(a_1|x_1,\lambda)P_1(a_2,a_3|x_2,x_3,\lambda)
\nonumber\\&& +p_2
\rho_2(\lambda)P_2(a_2|x_2,\lambda)P_2(a_1,a_3|x_1,x_3,\lambda)
\nonumber\\&& +p_3
\rho_3(\lambda)P_3(a_3|x_3,\lambda)P_3(a_1,a_2|x_1,x_2,\lambda)\big]\
.
 \label{Svet}
\end{eqnarray}
Here $P(a_2,a_3|x_2,x_3,\lambda)$, and the two terms like it, can
be $any$ probability distribution (it need not separate into
$P_1(a_2|x_2,\lambda)P_1(a_3|x_3,\lambda)$ \cite{note that}). The
main result of Svetlichny \cite{Svet} is to show that there are
quantum states (e.g. the Greenberger-Horne-Zeilinger state
$(1/\sqrt{2})(\ket{000} + \ket{111})$; see \cite{Pop} for a proof
that this is the optimal state for this purpose) with $P_{qm}(\vec
a| \vec x)$ such that no distribution $P_{Svet}(\vec a| \vec x)$
can be found such that $P_{qm}=P_{Svet}$ for all $\vec x$. This is
proved by introducing an inequality - called a ``Sveltichny
inequality''- which must be satisfied by correlations of the form
eq. (\ref{Svet}) but can be violated by quantum correlations. Thus
even allowing some non-local, ie. superluminal, classical
communication between pairs of parties, one cannot reproduce all
three party quantum correlations.

Refs. \cite{Seev, Col} extended the hybrid local/non-local model
of Svetlichny to the $m$-party setting. They allowed arbitrary
communication within disjoint subsets of the parties, but each
subset was independent of the settings and results of other
subsets. In Collins et al. these correlations were termed
``separable''. It was shown that the GHZ state has measurement
statistics which cannot be reproduced by these models. Sets of
correlations which are not separable will be called
``inseparable''.

In the present paper we will show that there are many inseparable
correlations which cannot reproduce  quantum correlations. More
precisely we will define a class of ``Partially Paired''
correlations $P_{PP}(\vec a| \vec x)$, which include the separable
correlations and some inseparable correlations, and categorize
them in terms of networks of communication. Using a generalised
Svetlichny inequality, taken from \cite{Col}, we show that this
class cannot reproduce all quantum correlations. We will also show
that the complement of this class, the ``Totally Paired''
correlations, $P_{TP}(\vec a| \vec x)$ maximally violate these
inequalities. These generalised Svetlichny inequalities therefore
cannot discriminate between models in $TP$ and quantum
correlations. The class $TP$, unlike $PP$, is thus a good
candidate for a classical description of all quantum correlations.

Our results are based on two advances; first we have formulated an
intuitive graphical means of classifying multi-particle
correlations allowing us to define $PP$ and $TP$; second we
developed a deeper understanding of the multi-particle Svetlichny
Inequality and the set of all similar inequalities. Both of these
advances promise results beyond the scope of this paper.

In the following we first introduce some of the basic conceptual
tools we shall use in our work (\ref{II},\ref{III}), we then
sketch our results in the four-party case (\ref{IV},\ref{V})
before providing full proofs for $m$-parties (\ref{VI},\ref{VII}).

\section{Geometry of the space of correlations} \label{II}

Consider $m$ parties, each of which receives an input $x_i$ and
produces an output $a_i$. The outputs can be correlated to the
inputs in an arbitrary way. Hence the most general way of
describing such a situation, independently of any underlying
physical model,
 is by a set of probability distributions $P(\vec a|\vec x)$.
The starting point of our investigation is to describe, in detail,
the geometry of the set of such probability distributions.

The set of probability distributions is characterised by the
normalisation conditions:
\begin{equation}
\sum_{\vec a} P(\vec a|\vec x)=1,
\label{normalisation}\end{equation} and the positivity conditions:
\begin{equation}
 P(\vec a|\vec x) \geq 0\ .
\label{positivity}\end{equation} Therefore the set of possible
probability distributions is a convex polytope. This polytope
belongs to the subspace defined by eq. (\ref{normalisation}). Its
facets are given by equality in  (\ref{positivity}).

It is useful to find  the extreme points of this polytope. These
are the probability distributions which saturate a maximum of the
positivity conditions, eq. (\ref{positivity}), while satisfying
the normalisation conditions, eq. (\ref{normalisation}). It is
easy to see that the extreme points are the probability
distributions such that, for each $\vec{x}$, there is a unique
$\vec{a}=\vec{a}(\vec{x})$ with $P(\vec{a}|\vec{x})=1$ (and
therefore if $\vec{a}\neq \vec{a}(\vec{x})$ then
$P(\vec{a}|\vec{x})=0$). Thus there is a one to one correspondence
between the set of extreme points and the functions
$\vec{a}(\vec{x})$ from the inputs to the outputs. Any particular
$\vec{a}(\vec{x})$ defines an extreme point. We call the extreme
points {\em deterministic models} since there is no randomness in
this case: the output is completely fixed by the input. We will
soon associate a graph with families of extreme points.

Subspaces of the space of all distributions satisfying
normalisation and positivity can be defined by taking all convex
combinations of a subset of the extreme points. This is a natural
construction because, if a physical model can produce certain
extreme points, then it can produce any convex combination of
these extreme points simply by randomly choosing which extreme
point to realize.

A first interesting example is the subspace defined by  {\em lhv}
models. The corresponding extreme points are of the form
$a_i(\vec{x})=a_i({x_i}$): the measurement results of party $i$
depend only on the settings of that party.

A second example  subspace is provided by the separable
correlations considered by Svetlichny and in \cite{Seev, Col}. The
corresponding extreme points can be characterised as follows. For
each extreme point there is a partition of the set of all parties
into two subsets, say $\{1,...,k\}$ and $\{k+1,...,m\}$, such that
$a_i(\vec{x})=a_i(x_1,...,x_k)$ if $i\in \{1,...,k\}$ and
$a_i(\vec{x})=a_i(x_{k+1},...,x_m)$ if $i \in \{k+1,...,m\}$. In
the former situation, party $i$ has results, $a_i$, independent of
the settings of the set $\{k+1,...,m\}$ and dependent on the
settings $(x_1,...,x_k)$.

Note that the formulation given by Svetlichny, see eq.
(\ref{Svet}), and by \cite{Seev,Col} may seem  more general than
this since in their formulation the outputs of any party in set
$\{1,...,k\}$ can depend on the inputs {\em and} outputs of all
parties $1,...,k$ whereas above we have only allowed the outputs
of any party in set $\{1,...,k\}$ to depend on the inputs of all
parties $1,...,k$.

Let us now show that this is not the case, and that the two
formulations are equivalent. For definiteness we will focus on the
three party case, eq. (\ref{Svet}), but the argument immediately
extends to an arbitrary number of parties. Consider the set of
probabilities $P_1(a_1|x_1 \lambda)$,
$P_1(a_2,a_3|x_2,x_3,\lambda)$ etc., appearing in equation
(\ref{Svet}). The key point  to note is that we can take these
probabilities to be deterministic strategies. Indeed we have just
argued that any  probability can be written as a convex
combination of extremal probabilities:
$P_1(a_2,a_3|x_2,x_3,\lambda)=\sum_\mu
p_\lambda(\mu)P_{\mu}^{ext}(a_2,a_3|x_2,x_3)$, where
$p_\lambda(\mu)$ is a probability distribution over $\mu$ and
where $P^{ext}_\mu(a_2,a_3|x_2,x_3)$ equals zero except if
$a_2=a_2(x_2,x_3)$ and $a_3=a_3(x_2,x_3)$. One can now suppose
that the variables, $\mu$, which specify the weighting of each
extremal probability, are included in the {\em lhv} variable
$\lambda$, i.e. the {\em lhv} tells the parties in each set what
deterministic strategy they must use. We have now proved our claim
since, in the case of extremal correlations, each output depends
only on the inputs of the parties, not on their outputs. Thus in
the case of separable correlations, letting the outputs of the
parties in each set depend only on the inputs of the parties in
their set is completely general: there is no need to also let them
depend on the outputs of the  parties in the set.

We now introduce new subspaces of the probability distributions
which form the basis for the present analysis.
\section{Classifying probability distributions using communication
  patterns}\label{secCommPattern} \label{III}

We will classify the extremal points by the settings that each
variable $a_i$ depends on. This dependence can be represented
using directed graphs. An arrow from party $i$ to $j$ means that
$a_j$ depends on the setting $x_i$. If there is no arrow from $i$
to $j$, $a_j$ is independent of $x_i$ (changing the value of $x_i$
only, leaves $a_j$ unaltered). The graph is directed as one might
have $a_i(x_j)$ but $a_j$ independent of $x_i$ (an arrow
$j\rightarrow i$ but no arrow $j\leftarrow i$). We call such a
graph a {\em communication pattern}.

Any such communication pattern can be associated naturally to a
model in which

i) Each party receives its input.

ii) If there is an arrow $i\rightarrow j$, $i$ sends its input to
$j$.

iii) The parties which receive arrows produce their measurement
results conditional on the list of inputs sent to them and their
own input.

Steps ii),iii) should be thought of as a single-shot mail
strategy: all of the settings are posted at the same time, along
the appropriate arrow, and they are received at the same time. On
receipt, the parties immediately generate their results: there is
no communication about the results obtained. It is now clear why
the arrows are directed: $i$ sending a letter to $j$ does not
imply that $j$ mails $i$. Note that this single-shot mailing has
to be superluminal if the measurements take place simultaneously
and at spatially separated locations.

Note that any particular graph represents many extreme points.
Indeed each extreme point corresponds to a unique function
$\vec{a}=\vec{a}(\vec{x})$ whereas each graph only determines the
variables the functions $\vec{a}$ depend on.

A formalisation of the above will prove useful. A given graph $G$
represents a set, $E_G$, of  extreme points. Each point is
identifiable with a different function $\vec{a}(\vec{x})$. These
 extreme points can be combined in convex combinations to make
different distributions $P(\vec{a}|\vec{x})$ for each $\vec{x}$.
The set of all such correlations produced by convex combinations
of the extremal points $E_G$ will be called $C_G$: the set of
correlations of type $G$. It is important to note that we define
the set $E_G$ as including the  extremal points represented by all
subgraphs of $G$. For example,  the graph $F$, in which all points
send arrows to all others (Fig. $1v$), represents the set of all
extreme points. Hence $C_F$ is the space of all correlations.

\begin{figure}[h!]
\includegraphics[angle = 0, width = 6cm, height = 11cm]{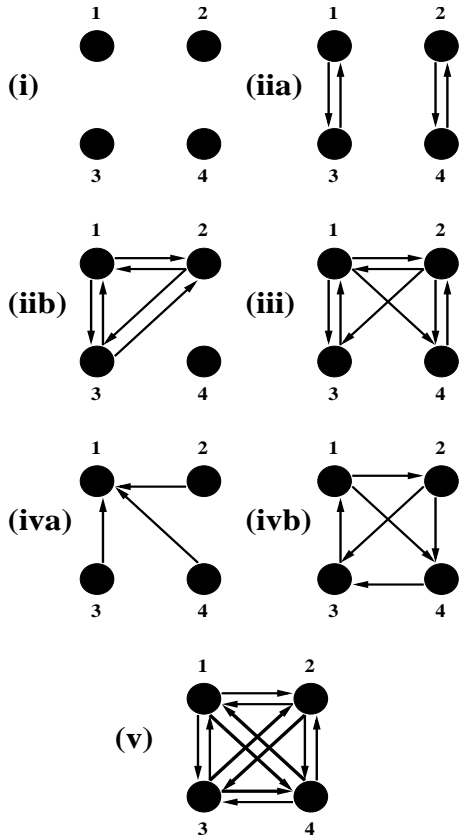}
\caption{\label{fig1}i) This graph represents  the set of extreme
points identifiable with {\em lhv} models, ie. $a_i(x_i)$. Bell
considered convex combinations of these points. ii a) and b) These
graphs represent functions $\vec{a}(\vec{x})$ with the structure
$a_1(x_1,\wedge,x_3,\wedge)$, $a_3(x_1,\wedge,x_3,\wedge)$,
$a_2(\wedge,x_2,\wedge,x_4)$, $a_4(\wedge,x_2,\wedge,x_4)$ and
$a_1(x_1,x_2,x_3,\wedge)$, $a_3(x_1,x_2,x_3,\wedge)$,
$a_2(x_1,x_2,x_3,\wedge)$, $a_4(\wedge,\wedge,\wedge,x_4)$
respectively. \cite{Col} considers correlations which are convex
combinations of
 extremal points represented by separable graphs of
this form and show that such models have $\langle S_4\rangle \leq
2$ (see Section \ref{IV}) \
 iii) This
graph represents functions $\vec{a}(\vec{x})$ which have
$a_1(x_1,x_2,x_3,\wedge)$, $a_3(x_1,x_2,x_3,\wedge)$,
$a_2(x_1,x_2,\wedge,x_4)$, $a_4(x_1,x_2,\wedge,x_4)$. Parties $3$
and $4$ are separated in the sense that no party receives arrows
from both. We will show that convex combinations of these extreme
points yield a value of $\langle S_4\rangle \leq 2$. We  call
graphs of type i)-iii) `Partially Paired'. \ iv) Examples of
graphs in which there is always a party which knows the settings
of any pair - we call these `Totally Paired'. Certain convex
combinations of the
 extremal points represented by these graphs can achieve
the algebraic maximum $\langle S_4\rangle =4$, as described in the
text. a) $a_1(x_1,x_2,x_3,x_4)$, $a_3(\wedge,\wedge,x_3,\wedge)$,
$a_2(\wedge,x_2,\wedge,\wedge)$, $a_4(\wedge,\wedge,\wedge,x_4)$
b) $a_1(x_1,\wedge,x_3,\wedge)$, $a_3(\wedge,x_2,x_3,x_4)$,
$a_2(x_1,x_2,\wedge,\wedge)$, $a_4(x_1,x_2,\wedge,x_4)$ \ v) This
graph represents the set of all extreme points and all possible
4-party graphs are its subgraphs.}
\end{figure}

Different models, such as {\em lhv} models \cite{B} or separable
models \cite{Svet,Seev,Col} can be associated with different
classes of graphs. This is illustrated in the case of 4 parties in
Fig. \ref{fig1}. The notation $a_1(x_1,\wedge,\wedge,x_4)$ is to
be read as `party $1$'s outcome is independent of the settings of
parties $2,3$ but dependent on the settings of $1,4$', ie. $a_1$
is unaltered by changes in $x_2,x_3$.
\section{Four Party Case: the Svetlichny inequality}
\label{4ptyC}\label{IV}

Our study of multi-particle non-locality is based on combining the
classification of correlations in terms of communication patterns
with the Svetlichny inequality \cite{Svet} and its generalisation
described in \cite{Seev,Col}. We now illustrate this connexion in
the case of 4 parties.

From now on we restrict ourselves to the case where each party's
input is a single bit $x_i\in\{0,1\}$ and each party's output is a
single bit $a_i\in\{0,1\}$. In order to make contact with earlier
work we introduce an alternative notation.

First of all it is useful to suppose that the outputs have values
$+1,-1$ instead of $0,+1$. In this case we denote the output of
party $i$ by $A_i$ with the correspondence $A_i=(-1)^{a_i}$.

Secondly we denote by a superscript $x_i$ on $A_i$ (i.e.
$A_i^{x_i}$) the value of the input of party $i$. This traditional
notation is good in the case of {\em lhv} models, since each
output $A_i$ is uniquely determined by its input. But it is
somewhat unnatural for other models where $A_i$ can depend on  the
inputs of all parties and not only on $x_i$. However the product
of four outputs such as $A_1^1A_2^1A_3^0A_4^0$ is well defined in
this notation  since all inputs are specified. The expression
$A_1^1A_2^1A_3^0A_4^0$ denotes the product of the outputs of
parties $1,2,3$ and $4$ given that the inputs are
$(x_1,x_2,x_3,x_4)=(1,1,0,0)$.

We will be interested in the Svetlichny polynomials which are
specific combinations of all products
$A_1^{x_1}A_2^{x_2}A_3^{x_3}A_4^{x_4}$. In the case of four
parties the Svetlichny polynomial is
\begin{eqnarray}
S_4&=&\frac{1}{4}
\Big[A_1^0A_2^0(-A_3^0A_4^0+A_3^0A_4^1+A_3^1A_4^0+A_3^1A_4^1)\nonumber
\\
 &&\;\; +A_1^1A_2^0(A_3^0A_4^0+A_3^0A_4^1+A_3^1A_4^0-A_3^1A_4^1) \nonumber
\\
&&\;\;
+A_1^0A_2^1(A_3^0A_4^0+A_3^0A_4^1+A_3^1A_4^0-A_3^1A_4^1)\nonumber
\\
&&\;\;
-A_1^1A_2^1(-A_3^0A_4^0+A_3^0A_4^1+A_3^1A_4^0+A_3^1A_4^1)\Big]\nonumber\\
&=& \sum_{\vec x} F_4(\vec x)
A_1^{x_1}A_2^{x_2}A_3^{x_3}A_4^{x_4}\ . \label{s4}
\end{eqnarray}

The expectation value of the Svetlichny polynomial is obtained by
taking the average over all possible inputs and outputs weighted
by the corresponding probabilities. Explicitly this can be written
as
\begin{equation}
\langle S_4 \rangle = \sum_{\vec x} F_4(\vec x) (-1)^{a_1}
(-1)^{a_2}(-1)^{a_3} (-1)^{a_4} P(\vec a|\vec x)
\end{equation}

The basis of our analysis will be to compare the maximum value of
the Svetlichny polynomial attained by different models. Note that
since $\langle S_4\rangle$ is a linear function of the
probabilities $P(\vec a|\vec x)$, its maximum value when the
$P(\vec a|\vec x)$ belong to a convex space will be attained on
the extreme points of this space. This is the main justification
for classifying correlations according to their extreme points:
Bell type expressions obtain their maximum value on the extreme
points.

It is easy to show that in the case of {\em lhv} models $\langle
S_4\rangle  \leq 2$. Collins et al. show that for separable
models, $\langle S_4\rangle\leq 2$.  And for certain quantum
states the value of $\langle S_4\rangle =2^{\frac{3}{2}}$
\cite{Pop}. We will show that much more general non-local models
than the separable models considered in \cite{Seev,Col} also have
$\langle S_4\rangle \leq 2$. Thus quantum mechanical states can
exhibit even more general types of non-locality than previously
anticipated.

\section{Four party case: Communication patterns and the Svetlichny
inequality}\label{V}

Collins et al. showed that the set of correlations of type Fig.
1i),iia),iib) describe statistics of `separable' physical models
that cannot simulate all quantum states. We will now show that,
despite being much more correlated than ii), no extreme point
represent by graph iii) has a larger maximum value for $S_4$.
Below is a sketch of a proof, details being saved for the
$m$-party setting.

Consider a deterministic setting, $\vec{a}=\vec{a}(\vec{x})$,
characterised by the graph iii) where
$a_1=a_1(x_1,x_2,x_3,\wedge)$, $a_3=a_3(x_1,x_2,x_3,\wedge)$,
$a_2=a_2(x_1,x_2,\wedge,x_4)$, $a_4=a_4(x_1,x_2,\wedge,x_4)$.
Noting the form of the $a_i$, the following term from (\ref{s4}),
\begin{equation}
\frac{1}{4}A_1^0A_2^0(-A_3^0A_4^0+A_3^0A_4^1+A_3^1A_4^0+A_3^1A_4^1),
\label{firstterm}
\end{equation}
can be rewritten as:
\begin{eqnarray}
\frac{1}{4}&&\Big[-(-1)^{a_1(0,0,0,\wedge)+a_2(0,0,\wedge,0) +
a_3(0,0,0,\wedge)
+a_4(0,0,\wedge,0)}\nonumber\\
+&&(-1)^{a_1(0,0,0,\wedge)+a_2(0,0,\wedge,1) + a_3(0,0,0,\wedge)
+a_4(0,0,\wedge,1)}\nonumber\\
+&&(-1)^{a_1(0,0,1,\wedge)+a_2(0,0,\wedge,0) + a_3(0,0,1,\wedge)
+a_4(0,0,\wedge,0)}\nonumber\\
+&&(-1)^{a_1(0,0,1,\wedge)+a_2(0,0,\wedge,1) + a_3(0,0,1,\wedge)
+a_4(0,0,\wedge,1)}\Big].
\end{eqnarray}
One can now find the maximum value of this expression. Defining
$a_1(0,0,x_3,\wedge)+a_3(0,0,x_3,\wedge)\;\; \rm mod \rm
\;\;2=\beta_1(x_3)$ and
$a_2(0,0,\wedge,x_4)+a_4(0,0,\wedge,x_4)\;\; \rm mod \rm
\;\;2=\beta_2(x_4)$ (this approach will be reused later) this
expression simplifies to:
\begin{eqnarray}
&&\frac{1}{4}\Big[-(-1)^{\beta_1(0)+\beta_2(0)}+(-1)^{\beta_1(0)+\beta_2(1)}\nonumber\\&&
+(-1)^{\beta_1(1)+\beta_2(0)}+(-1)^{\beta_1(1)+\beta_2(1)}\Big].
\end{eqnarray}
One now notes that, whatever the value of functions $\beta_i$,
this expression is $\leq\frac{1}{2}$. Similarly each of the three
other terms in $S_4$:
\begin{eqnarray}
 && \frac{1}{4}A_1^1A_2^0(A_3^0A_4^0+A_3^0A_4^1+A_3^1A_4^0-A_3^1A_4^1), \nonumber
\\
&&
\frac{1}{4}A_1^0A_2^1(A_3^0A_4^0+A_3^0A_4^1+A_3^1A_4^0-A_3^1A_4^1),\nonumber
\\
&
-&\frac{1}{4}A_1^1A_2^1(-A_3^0A_4^0+A_3^0A_4^1+A_3^1A_4^0+A_3^1A_4^1),
\end{eqnarray}
 can have a maximum value of $\frac{1}{2}$ and  can
do so simultaneously: the maximum value of $S_4$ is thus $2$. By
convexity, even a probabilistic mix of all strategies of the form
i),iia,b),iii) still cannot exceed $S_4=2$.

Thus very strongly correlated graphs, representing inseparable
probability distributions, can still fail to exceed $S_4=2$. If
two parties in a graph are   separated from each other, or
`unpaired', such that no party knows the settings of both, then,
no matter how correlated the remainder of the state,  $S_4$ cannot
exceed $2$.

  This observation motivates the definition of
``Partially Paired'' graphs given in Fig.2 and in definition 1
below. Indeed it will be shown that, despite being highly
connected, convex combinations of extremal points defined by
``Partially Paired'' graphs  cannot replicate all quantum
statistics.
\begin{figure}[h!]
\includegraphics[angle = 0, width = 6cm, height =3.75cm]{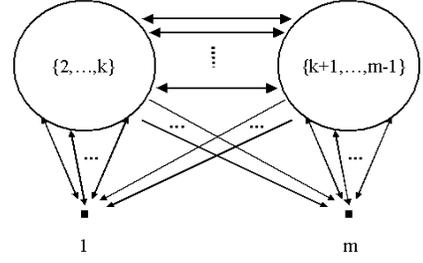}
\caption{\label{fig2} Graphical representation of the most general
Partially Paired  (PP) $m$-party graph. One circle contains
parties $\{2,...,k\}$ and the other circle contains parties
$\{k+1,...,m-1\}$. All parties within a circle can send and
receive arrows from all other parties within the same circle. Each
party in a circle can send and receive arrows to all parties in
the other circle.  Party 1 can send and receives arrows from
parties $\{2,...,k\}$. Party 1 can receive arrows from (but does
not send arrows to) parties $\{k+1,...,m-1\}$. Party $m$ can
receive and send arrows to parties $\{k+1,...,m-1\}$. Party $m$
can receive arrows from (but does not send arrows to) parties
$\{2,...,k\}$. There are no arrows between parties 1 and $m$. With
these restrictions there is no party that receives arrows both
from party 1 and party $m$. The subgraph composed only of the
arrows emanating from parties 1 and $m$ is therefore separable.
The graph thus belongs to the PP class.}\end{figure}

By contrast, graphs in which, for any given pair of settings,
there is always a party with results dependent on that pair (such
as Fig. 1iv a,b)), can represent  extreme points which reach the
maximum value of $S_4$, namely $4$. It is straightforward to show
that the graph of Fig. 1iva) with $a_1(x_1,x_2,x_3,x_4)$ can
represent a function $\vec{a}(\vec{x})$ which will obtain the
algebraic maximum. The graph in Fig. 1ivb), despite having
apparently less communication than iii), also has extremal points
which reach this maximum. Recalling, for graph ivb) that,
$a_1(x_1,\wedge,x_3,\wedge),a_3(\wedge,x_2,x_3,x_4),$
$a_2(x_1,x_2,\wedge,\wedge),a_4(x_1,x_2,\wedge,x_4)$,
(\ref{firstterm}) becomes:
\begin{eqnarray}
&&\frac{1}{4}\Big[-(-1)^{a_1(0,\wedge,0,\wedge)+a_2(0,0,\wedge\wedge)
+ a_3(\wedge,0,0,0) +a_4(0,0,\wedge,0)}\nonumber\\&&
+(-1)^{a_1(0,\wedge,0,\wedge)+a_2(0,0,\wedge,\wedge) +
a_3(\wedge,0,0,1) +a_4(0,0,\wedge,1)}\nonumber\\
&& +(-1)^{a_1(0,\wedge,1,\wedge)+a_2(0,0,\wedge,\wedge) +
a_3(\wedge,0,1,0) +a_4(0,0,\wedge,0)}\nonumber\\
&& +(-1)^{a_1(0,\wedge,1,\wedge)+a_2(0,0,\wedge,\wedge) +
a_3(\wedge,0,1,1) +a_4(0,0,\wedge,1)}\Big]
\end{eqnarray}
If we set $a_3(\wedge,0,1,0)=a_4(0,0,\wedge,0)=1$ and all of the
other terms above equal to zero the expression takes value $1$. It
is relatively easy, to find, by inspection a set of $a_i$ such
that $S_4$ reaches its algebraic maximum of $4$. Indeed defining
$a_1(x_1,x_3)=x_1x_3+x_1;\;\;a_2(x_1,x_2)=x_1x_2+x_2;\;\;
a_3(x_2,x_3,x_4)=x_2x_3+x_3x_4+x_3;\;\;a_4(x_1,x_2,x_4)=x_1x_4
+x_2x_4 +x_4+1$ obtains the maximum value for $S_4$. In this case
it may be seen that:
\begin{equation} \sum^4_{i=1} a_i(\vec{x}) = \sum_{i<j}^4 x_ix_j +
  \sum_{i=1}^4x_i +1. \label{teaser}\end{equation} This
form will prove significant later.

 A deterministic strategy with
functional form represented by Fig. 1iii) is non-local: it can be
pictured as requiring faster than light correlations. Nonetheless
it cannot always replicate quantum statistics as we have described
above. Conversely, as we described, extremal points represented by
graphs like iva,b) can reach the algebraic maximum of $S_4$,
namely 4.

Note that individual extremal points represented by graphs like
iva,b) are signalling: by looking at the outputs of one party one
can learn about the inputs of other parties. We will show below,
however, that there exist convex combinations of extremal points
represented by graphs like iva,b) which are no-signalling and
which reach the algebraic maximum of $S_4$. This is an important
remark since special relativity only permits no-signalling
correlations.

Before moving to the $m$-party generalisation, we review: we have
identified two kinds of four party correlations. The first are
correlations represented by graphs i)-iii) and the second
represented \it only \rm by graphs like iv). The former reach a
value of $\langle S_4\rangle \leq 2$ the later can reach $S_4=4$.
For all quantum states $\langle S_4\rangle  \leq 2\sqrt{2}$.

 \section{The $m$-party Svetlichny Polynomial}
 \label{mptyC}\label{VI}

Having sketched proofs for the four-party Svetlichny polynomial,
$S_4$, we can provide general proofs for the $m$-party setting
with polynomials $S_m$. We first recall the definition of $S_m$,
as formulated in  \cite{Col}.

We again consider the situation where each party's input
$x_i\in\{0,1\}$ is a single bit and each party's output
$a_i\in\{0,1\}$ is a single bit. We will use the notation
introduced at the beginning of section \ref{4ptyC}. As in
\cite{Col} we construct $S_m$ from the Mermin polynomials. (The
Mermin inequality is just one of the correlation inequalities
described in \cite{Wolf,Zuk} where  all correlation inequalities
for $n$ parties, each having 2 inputs and 2 outputs, were
characterised). Let the two party Mermin polynomial be
\begin{equation}
M_2=\frac{1}{2}(A^0_1A^0_2+A^1_1A^0_2+A^0_1A^1_2-A^1_1A^1_2),
\label{c}
\end{equation}
and  define the new notation:
\begin{equation}
M_m\equiv \sum_{\vec x}F_m(\vec
x)A_1^{x_1}A_2^{x_2}...A_{m}^{x_m}, \label{cc}
\end{equation}
where $\vec x = (x_1,...,x_m)$ and also
\begin{eqnarray}
M'_m &\equiv& \sum_{\vec x}F_m(\vec x)A_1^{\overline
x_1}A_2^{\overline x_2}...A_{m}^{\overline x_m}\nonumber\\
&=&\sum_{\vec x}F_m(\vec {\overline
x})A_1^{x_1}A_2^{x_2}...A_{m}^{x_m} \label{ccc}
\end{eqnarray}
where $\overline x_i=x_i+1 \mod 2$. $M_m$ is generated from
$M_{m-1}$ by the recursion relation:
\begin{equation}
M_m=\frac{1}{2}\Big[M_{m-1}(A_m^0+A_m^1)+M'_{m-1}(A^0_m-A^1_m)\Big].
\label{d}
\end{equation}
Using this twice yields:
\begin{equation}
M_{m+2}=\frac{1}{2}\Big[M_{m}(M_{2}+M'_{2})+M'_{m}(M_{2}-M'_{2})\Big]
.
 \label{ee}
\end{equation}
The recursion relation (\ref{ee}) can be written in terms of
$F_{m}$ as:
\begin{eqnarray}
&&F_{m+2}(\vec x)=\nonumber \\&& \quad\quad
\frac{1}{2}\bigg[F_m(\vec x)\Big( F_{2}(x_{m+1},
x_{m+2})+F_2(\overline x_{m+1}, \overline x_{m+2})\Big)
\nonumber\\
&&\quad\quad  +F_m(\vec {\overline x})\Big(F_2(x_{m+1},
x_{m+2})-F_2(\overline x_{m+1}, \overline x_{m+2})\Big)\bigg]\ . \nonumber\\
\label{h}
\end{eqnarray}
Following \cite{Col} we define the Svetlichny polynomials as
\begin{eqnarray}
S_m=\begin{cases} M_m,& \text{if $m$ is even},\label{a}\\
\frac{1}{2}(M_m+M'_m),& \text{if $m$ is odd}.
\end{cases}
\end{eqnarray}
We also define
\begin{equation}
S_m=\sum_{\vec x}\mu_m(\vec x) A^{x_1}_1...A_m^{x_m}, \label{bb}
\end{equation}
which implies:
\begin{eqnarray}
\mu_m(\vec x)=\begin{cases}F_m(\vec x),& \text{if $m$ is even}\\
\frac{1}{2}\Big[F_m(\vec x) + F_m( {\vec
  {\overline x}})\Big],& \text{if $m$ is odd}.
  \label{23}\end{cases}
\end{eqnarray}

In order to have a useful characterisation of $S_m$ we will obtain
an explicit expression for $\mu_m$ and $F_m$:

{\bf Lemma 1} {\em Let $q=\lceil m/2\rceil$ (where $\lceil \rceil$
indicates rounding up to the next nearest integer). Then
\begin{equation}
\mu_m(\vec x )=\frac{1}{2^q}(-1)^{\left[\sum^m_{i<j}x_ix_j +
(q+1)\sum_{i=1}^mx_i + (q^2-q)/2\right]}, \label{i}
\end{equation}
and:
\begin{equation}
F_{2k+1}(\vec x)=\frac{1}{2}F_{2k}(x_1,\ldots,x_{2k})
\left(1+(-1)^{\sum_{i=1}^{2k} x_i + x_{2k+1} + k}\right)
\label{ii}\ .
\end{equation}
}

As an example, note that in the four party, case one finds:
\begin{equation}
\mu_4(\vec x)=\frac{1}{4}(-1)^{\left[\sum_{i<j}^4x_ix_j
+\sum_{i=1}^4x_i+1\right]}, \label{ij}
\end{equation}
which (combined with (\ref{bb})) reproduces eq. (\ref{s4}). Note
that the exponent in (\ref{ij}) is identical to (\ref{teaser}).

We now turn to the proof of Lemma 1.

\subsection*{Proof of Lemma 1.}

\subsubsection*{Proof of Eq. (\ref{i}) for m=2k,  k integer.}

In this case $\mu_{{2k}}=F_{{2k}}$. One easily checks that Lemma 1
is true for $k=1$ when
\begin{eqnarray}
F_2(x_1,x_2)&=&\frac{1}{2}(-1)^{x_{1}x_{2}},\nonumber\\
F_2(\overline x_1,\overline x_2)&=&\frac{1}{2}(-1)^{x_{1}x_{2} +
x_{1}+x_{2}+1}\ . \nonumber\end{eqnarray} We now proceed by
induction. We suppose Lemma 1 is true for $k$ and will show it is
true for $k+1$.
 From eq. (\ref{i}):
\begin{equation}
F_{2k}(\vec{\overline
x})=\frac{1}{2^{k}}(-1)^{[\sum^{2k}_{i<j}(x_i+1)(x_j+1) +
(k+1)\sum_{i=1}^{2k}(x_i+1) + (k^2-k)/2]} \label{iii}
\end{equation}
where $F_{2k}(\vec{ x})=F_{2k}(x_1,...,x_{2k})$ and $q=k$. Eq.
(\ref{iii}) can be written as:
\begin{equation}
F_{2k}(\vec{\overline x})=
\frac{1}{2^k}(-1)^{[\sum^{2k}_{i<j}x_ix_j + k\sum_{i=1}^{2k}x_i +
(k^2+k)/2]},
\end{equation}
using the identities
\begin{equation}\sum_{i<j}^{2k}\overline x_i\overline x_j=\sum_{i<j}^{2k}x_ix_j +\sum_i^{2k}x_i +k\;\; \text{ mod}\;\;2,\end{equation} and
$(-1)^{2\alpha}=1$, $(-1)^{-\alpha}=(-1)^{\alpha}$ for $\alpha$
integer. Inserting $F_2(x_{2k+1},x_{2k+2})$, $F_2(\overline
x_{2k+1},\overline x_{2k+2})$, $F_{2k}(\vec{\overline x})$,
$F_{2k}(\vec{x})$ into (\ref{h}) and noting that,
\begin{equation}
F_{2k}(\vec{x})=(-1)^{\sum^{2k}_{i=1}x_i+k} F_{2k}(\vec{\overline
x}), \label{j}
\end{equation}
 the righthand side of (\ref{h}) becomes:
 \begin{eqnarray}
&&F_{2k+2}(\vec{x})=\frac{1}{4}F_{2k}(\vec{\overline
x})(-1)^{x_{2k+1}x_{2k+2}}(1+(-1)^{x_{2k+1}+x_{2k+2}}\nonumber\\&&
+ (-1)^{\sum_{i=1}^{2k}x_i
+k}+(-1)^{x_{{2k}+1}+x_{{2k}+2}+\sum_{i=1}^{2k}x_i
+k +1})\nonumber\\
&&=\frac{1}{2}F_{2k}(\vec{\overline
x})(-1)^{(x_{{2k}+1}+x_{{2k}+2})(\sum_{i=1}^{2k}x_i +k)+
x_{{2k}+1}x_{{2k}+2}},\label{iv}
\end{eqnarray}
where we have used the identity:
\begin{equation}
(-1)^{(ab)}=\frac{1}{2}(1+(-1)^b+(-1)^a+(-1)^{a+b+1}), \label{k}
\end{equation}
for $a,b$ integer. Eq. (\ref{iv}) can be rewritten as:
\begin{eqnarray}
&&F_{{2k}+2}(\vec{x})=\frac{1}{2^{k+1}}(-1)^{\sum_{i<j}^{2k}x_ix_j
+
(x_{{2k}+1}+x_{{2k}+2})(\sum_{i=1}^{2k}x_i)}\times\nonumber\\&&(-1)^{x_{{2k}+1}x_{{2k}+2}+(k+2)\sum_{i=1}^{2k}x_i+(k+2)(x_{{2k}+1}+x_{{2k}+2})+(k^2+k)/2}.\nonumber\\
&&
\end{eqnarray}
One can readily check that this coincides with Eq. (\ref{i}) for
$m=2k+2,\;\;q=k+1$. Eq. (\ref{i}) thus satisfies the recursion
relation (\ref{h}) for ${2k}$ even. $\Box$

\subsubsection*{Proof of Eq. (\ref{ii}). }

 Eq. (\ref{d}) can be rewritten as:
\begin{eqnarray}
M_{{2k}+1}=&&\frac{1}{2}\Big[M_{2k}\sum^1_{x_{{2k}+1}=0}A^{x_{2k}+1}_{{2k}+1}\nonumber\\&&+M'_{2k}\sum_{x_{{2k}+1}=0}^1(-1)^{x_{{2k}+1}}A^{x_{{2k}+1}}_{{2k}+1}\Big].
\end{eqnarray}
Using (\ref{cc}), this is equivalent to the relation:
\begin{equation}
F_{{2k}+1}(\vec x)=\frac{1}{2}\Big[F_{2k}(\vec
x)+F_{2k}(\vec{\overline x})(-1)^{x_{{2k}+1}}\Big]. \label{ex}
\end{equation}
Substituting eq. (\ref{j}) into eq. (\ref{ex}) one recovers eq.
(\ref{ii}). $\Box$

\subsubsection*{Proof of Eq. (\ref{i}) for m=2k+1, k integer.}

Inserting equations (\ref{ex}) and  (\ref{j}) into eq. (\ref{23}),
for $m$ odd, yields:
\begin{eqnarray}
&&\mu_{{2k}+1}(\vec x)=\frac{1}{4}F_{2k}(\vec{\overline
x})(1+(-1)^{\sum_{i=1}^{2k}(x_i+1) + x_{2k} +k+1}+\nonumber
\\&&(-1)^{\sum^{2k}_{i=1}x_i +k}(1+(-1)^{\sum^{2k}_{i=1}x_i
+k+x_{{2k}+1}})).
\end{eqnarray}
This becomes, by identity (\ref{k}),
\begin{equation}
\mu_{{2k}+1}(\vec x)=\frac{1}{2}F_{{2k}}(\vec{\overline
x})(-1)^{x_{{2k}+1}(\sum^{{2k}}_{i=1}x_i+k)}.\label{l}
\end{equation}
This can be rewritten as:
\begin{eqnarray}
\mu_{{2k}+1}(\vec
x)=&&\frac{1}{2^{k+1}}(-1)^{\sum_{i<j}^{2k}x_ix_j +
x_{{2k}+1}\sum^{2k}_{i=1}x_i}\times\nonumber\\&&
(-1)^{(k+2)(\sum_{i=1}^{2k}x_i + x_{{2k}+1})+(k^2+k)/2}
\end{eqnarray}
which coincides with (\ref{i}). $\Box$

\section{The Classes PP and TP} \label{VII}

The central result of this article is to obtain bounds on the
maximum value of $S_m$ attainable in different models. Let us
recall what is already known in this respect:

{\bf Theorem 1:

[SS2002\cite{Seev}],[CGPRS2002\cite{Col}],[MPR2002\cite{Pop}]}
{\em
\begin{enumerate}
\item Local hidden variable models and separable models satisfy
the Svetlichny inequality
\begin{equation}
\langle S_m\rangle_{\text{lhv, separable models}}\leq 2^{m-\lceil
m/2 \rceil - 1}\ .
\end{equation}
\item The maximum value of $S_m$ attainable by quantum mechanics
(reached by carrying out measurements on the GHZ state) is
\begin{equation}
\langle S_m \rangle_{\text{quantum mechanics}} \leq 2^{m-\lceil
m/2 \rceil -1/2}\ .
\end{equation}
\item The algebraic maximum of $S_m$ (obtained by  taking
 $A_1^{x_1}...A_m^{x_m}=\mu(\vec
x_m)/|\mu(\vec x_m)|$ and using eq. (\ref{i})) is
\begin{equation}
S_m[alg]=2^{m-\lceil m/2 \rceil}\ .
\end{equation}
\end{enumerate}
}

We now go back to the classification of extreme points in terms of
communication patterns introduced in section \ref{III}. We define
two classes of graph and formulate two theorems about them.

{\bf Definition 1: Partially Paired (PP) Graphs (see Fig. 2).} A
communication pattern represented by a directed graph in which
there exist two (or more) parties $i,j$ such that there is no
party with results dependent on $x_i,x_j$. Graphically, this
definition can be rephrased as: Take the subgraph composed only of
the vertices receiving arrows originating from $i$ and $j$ and the
arrows originating from $i$ and $j$. This graph is separable: it
can be split into two
 disconnected graphs one including vertex $i$ and the other $j$.

Examples are Fig. 1i-iii) since, in these graphs, there is always
a pair of vertices $i,j$  that neither send arrows to each other,
nor both send to the same party.

{\bf Definition 2: Partially Paired Correlations.} These are the
correlations that can be written as convex combinations of the
extremal correlations whose associated graph is partially paired.
These correlations form the set $C_{PP}$:
$$
C_{PP} = \cup_{G\in PP} C_G
$$
 (where PP is the set of all PP graphs). Note that $C_{PP}$ is the space of correlations associated with
all possible PP graphs, not just a single PP graph $G_{PP}$.

One of our main results is:

{\bf Theorem 2.}  {\em All Partially Paired correlations (ie. all
  correlations in the set $C_{PP}$)  satisfy the
multi-party Svetlichny inequality $S_m\leq 2^{m-q-1}$.}

Note that the Svetlichny inequality is violated by some quantum
states, see Theorem 1. Thus PP correlations cannot reproduce all
quantum correlations.

The complementary class to PP graphs are the Totally Paired
graphs:

{\bf  Definition 3: Totally Paired (TP) Graphs.}  Any graph which
is not $PP$. Graphically this can be rephrased as: for any two
parties $i,j$ there always exists a party $k$ such that $k$
receives arrows originating from $i$ and $j$ ($k$ could coincide
with $i$ or $j$).

Examples are Fig. 1iv,v) (see \cite{Gutin04} for a graph theoretic
analysis of TP graphs).

The definition of TP graphs will allow us to prove the complement
of Theorem 2. Namely we will show that for any TP graph $G_{TP}$,
there exist correlations whose associated graph is $G_{TP}$ and
which maximally violate the Svetlichny inequality.

But it should be noted that - except in the case of {\em lhv}
models - an extreme point is a deterministic signalling strategy:
one party's results provides information about the settings of
other parties. Thus one could argue that such extreme points are
unphysical and cannot reproduce the predictions of causal
theories, such as quantum mechanics, which do not allow
signalling. But we will show that different  strategies, with the
same associated graph $G_{TP}$, can be combined to produce
no-signalling correlations while continuing to maximally violate
the Svetlichny inequality. Thus maximal violation of the
Svetlichny inequality by correlations in $C_{G_{TP}}$, where
$G_{TP}$ is an arbitrary $TP$ graph, is compatible with causality.
These results are summarized in:

{\bf Theorem 3.} {\em For any Totally Paired communication graph
$G_{TP}$, there exist correlations  in the set $C_{G_{TP}}$ (the
set of correlations obtained by convex combinations of the
extremal points $E_{G_{TP}}$ whose associated graph is $G_{TP}$)
which both attain the algebraic maximum of the Svetlichny
polynomial and are no-signalling.}

Note that $C_{G_{TP}}$ is the set of correlations associated with
a single TP graph, $G_{TP}$.

\subsection*{Proof of Theorem 2}
The Svetlichny inequalities can be written in the form
\begin{eqnarray}
\langle S_m\rangle=\sum_{\vec x, \vec a} \mu_m(\vec
x)(-1)^{\sum_ia_i}P(\vec a| \vec x)\leq 2^{m-q-1} \label{z}\ .
\end{eqnarray}
In the class $PP$ there always exists at least one pair of
settings, say $x_1,x_m$, such that no party's outcome is dependent
on both. As in Fig. 2, the $m$ parties can be divided into the set
$\{1,...,k\}$ dependent on $x_1$ (and not $x_m$) and
$\{k+1,...,m\}$ dependent on $x_m$ (and not $x_1$). Defining
$\sum_{i=1}^ka_i\mod 2\equiv\beta_1$, $\sum_{i=k+1}^ma_i\mod
2\equiv\beta_2$, and rewriting the left hand side of (\ref{z})
using (\ref{i}):
\begin{small}
\begin{eqnarray}
&\langle S_m\rangle_{PP} = \quad\quad\quad& \nonumber\\&
\frac{1}{2^q} \sum_{x_2,...,x_{m-1}} (-1)^{\sum_{i<j,i\neq
1}^{m-1}x_ix_j + (q+1)\sum^{m-1}_{i=2}x_i+(q^2-q)/2}&
\nonumber\\&\times \bigg(\sum_{x_1,x_m,\vec
a}(-1)^{x_1x_m+(q+1+\sum_{i=2}^{m-1}x_i)(x_1+x_m)}&\nonumber\\&\times
(-1)^{\beta_1+\beta_2}P(\vec a|\vec x)\bigg)& \label{y}
\end{eqnarray}
\end{small}
By the definition of $\beta_1,\beta_2$,
\begin{equation}\sum_{x_1,x_m,\vec
a}(-1)^{x_1x_m+(q+1+\sum_{i=2}^{m-1}x_i)(x_1+x_m)}(-1)^{\beta_1+\beta_2}P(\vec
a|\vec x),\end{equation} has the same form as a $CHSH$ expression
\cite{CHSH} and thus has a modulus $\leq 2$. Substituting this
into (\ref{y}) yields $\langle S_m\rangle_{PP} \leq 2^{m-q-1}$.
$\Box$

Thus, at least insofar as the Svetlichny inequality is concerned,
the set $C_{PP}$ is only as strong as its subset, the separable
correlations considered in \cite{Seev,Col}.

\subsection*{Proof of Theorem 3.}

Any extreme point in the set of correlations, ie. any
deterministic scenario, has $\vec a = \vec a(\vec x)$ and so
(\ref{z}) becomes
\begin{equation}\langle S_m\rangle =\sum_{\vec x} \mu_m(\vec x)(-1)^{\sum_ia_i(\vec x)}\label{zz}
\end{equation}
In order to reach the algebraic maximum of $S_m$ we require
$(-1)^{\sum_ia_i(\vec x)} = \mu(\vec x)/|\mu(\vec x)|$. From Lemma
1, this means that the algebraic maximum is attained if
\begin{equation}
\sum^m_{i=1}a_{i}=\sum^m_{i<j}x_ix_j + (q+1)\sum_{i=1}^mx_i +
(q^2-q)/2 .\label{zzz}\end{equation}

Let us show that for any Totally Paired graph $G_{TP}$, there is
an extreme point whose associated graph is $G_{TP}$ and which
satisfies eq. (\ref{zzz}). To see this note the following:
\begin{enumerate}
\item $a_i(\vec x)$ is a function of $x_j$ only when there is an
arrow originating at vertex $j$ and ending at vertex $i$. (In
addition $a_i$ can always depend on $x_i$). \item For any pair of
vertices $i,j$ in a $TP$ graph there is always a vertex $k$ which
either receives edges from both members of the pair or is itself a
member of the pair (ie. $k$ can coincide with either $i$ or $j$).
The output of this vertex $a_k(\vec x)$ can thus be equal to any
function of $x_i$ and $x_j$ (Plus possibly other functions
depending on the other arrows leading into vertex $k$). By an
appropriate choice of these functions we can reproduce the term
$\sum_{i<j}^m x_i x_j$ on the right hand side of  eq. (\ref{zzz}).
\item The output of any vertex $i$ can depend on the input to
vertex $i$. Hence the output of vertex $i$ can contain a term
equal to $a_i(x_i)=c x_i + d$. By combining these terms we can
reproduce the term $(q+1)\sum_{i=1}^mx_i + (q^2-q)/2$.
\end{enumerate}
By combining points 2 and 3 above, we can satisfy eq. (\ref{zzz}).

Thus certain extreme points in $E_{G_{TP}}$ (where $G_{TP}$ is any
$TP$ graph) define functions $\vec a(\vec x)$ such that
(\ref{zzz}) holds. These reach the algebraic maximum of the
multi-party Svetlichny polynomials. Convex combinations of these
will also obtain the maximum. This is why (\ref{teaser}) has the
same form as the exponent of (\ref{ij}).

We now turn to the second part of Theorem 3. We will show that
different strategies, with the same associated graph, $G_{TP}$,
can be combined to produce no-signalling correlations while
continuing to maximally violate the Svetlichny inequality. Let us
first recall that by no-signalling correlations we mean
correlations $P(\vec a|\vec x)$ such that one subset of parties,
say parties $1,\ldots, k$, cannot communicate to the other parties
$k+1,\ldots,m$ by changing the settings of their measurement
device. Mathematically this is expressed by the condition that for
all $a_{k+1},\ldots, a_m$
\begin{equation}
\sum_{a_1,\ldots, a_k} P(\vec a | \vec x) = P (a_{k+1},\ldots,a_m
| \vec x) \label{NS}\end{equation} where the right hand side is
independent of $x_1,\ldots, x_k$.

Let us now consider a particular graph $G_{TP}$ in the set $TP$.
To this graph we can associate at least one deterministic strategy
$\vec a\,^0 (\vec x)$ such that eq. (45) holds. This deterministic
strategy is necessarily signalling, ie. eq. (\ref{NS}) is not
independent of $x_1,\ldots, x_k$. The first step in proving the
second part of Theorem 3 is to note that $\vec a \,^0 ( \vec x)$
is not the only deterministic strategy which has $G_{TP}$ as its
associated graph and which obeys eq. (45). In fact from $\vec a
\,^0 ( \vec x)$  we can easily construct a set of $2^{m-1}$
deterministic strategies that all have $G_{TP}$ as their
associated graph and obey eq. (45). To this end define the $m$
component vectors, $\vec{b}^{\mu}\in \{ 0, 1\}^m$ with the
property $\sum_i {b^{\mu}_i}\; \rm mod\;2 =0$. There are $2^{m-1}$
such vectors, $\vec{b}^{1},\vec{b}^{2}...\vec{b}\,^{2^{m- 1}}$.
Now we define $\vec{a}^{\mu}(\vec x)=\vec{a}\,^0 (\vec
x)+\vec{b}^{\mu}$. Note that $\sum_i a^{\mu}_i\; \rm mod\;2$ =$
\sum_i a^0_i\; \rm mod\;2$, hence eq. (45) holds for all
deterministic strategies $\vec{a}^{\mu}(\vec x)$.

The second step in proving the second part of Theorem 3 is to note
that, while staying constrained by the graph $G_{TP}$, the parties
need not use a deterministic strategy. Instead, before the
protocol starts, they can choose one value of $\mu$ at random,
according to some probability distribution $p(\mu) \geq 0$,
$\sum_\mu p(\mu) = 1$. They then carry out the deterministic
strategy $\vec a^\mu(\vec x)$. Since $\mu$ is chosen at random the
resulting correlations have the form
\begin{equation}
P(\vec a|\vec x) = \sum_\mu p(\mu) P^\mu (\vec a | \vec x)
\label{Pmu}\end{equation} where $P^\mu (\vec a | \vec x)$ are the
correlations obtained by using the deterministic strategy $\vec
a^\mu$. Thus the parties have formed a convex combination of
deterministic strategies, all with the same associated graph
$G_{TP}$.

Let us now show that if $p(\mu) = 2^{-(m-1)}$ is the uniform
distribution, then the correlations defined by eq. (\ref{Pmu}) are
no-signalling. This follows from the fact that the correlations
associated with $\vec a^\mu$ have the form \begin{small}$P^\mu
(\vec a | \vec x) = \delta\big(a_1- \small(a_1^0(\vec x) +
b^\mu_1\small)\big)\delta\big(a_2- \small(a_2^0(\vec x) +
b^\mu_2\small)\big)\ldots \delta\big(a_m-( a_m^0(\vec x) +
b^\mu_m)\big)$\end{small}. We can now show that
$P(a_{k+1},\ldots,a_m|\vec x)$ is independent of $x_1,...x_{k}$:
\begin{small}\begin{eqnarray} &&P(a_{k+1},\ldots,a_m|\vec x)= \frac{1}{2^{m-1}}\sum_{a_1,\ldots,a_k} \sum_\mu  P^\mu
(\vec a | \vec x)\nonumber\\\!\!\!\!&=&\!\!\!\!
\frac{1}{2^{m-1}}\!\!\!\!\sum_{a_1,...,a_k}\!\!\!\! \sum_{\;\;\mu}
\delta\big(a_1^{ }\!\!-\!\!( a_1^0(\vec x) + b^{\mu}_1)\big)...
\delta\big(a_m^{ }\!\!-\!\!
(a_m^0(\vec x) + b^{\mu}_m)\big) \nonumber\\
\!\!\!\!&=&\!\!\!\!\frac{1}{2^{m-1}}\sum_{\mu}
\Big(\delta\big(a_{k+1}^{ }\!\!-\!\!(a_{k+1}^0(\vec
x)+b^{\mu}_{k+1})\big) ...\delta\big(a_{m}^{
}\!\!-\!\!(a_{m}^0(\vec
x)+b^{\mu}_{m})\big)\nonumber\\&\;\;\;\;\;&\times\;\;
\sum_{a_1,\ldots,a_k}\delta\big(a_1\!\!-\!\!(a_1^0(\vec x) +
b^{\mu}_1)\big)\ldots \delta\big(a_k\!\!-\!\!(
a_k^0(\vec x) + b^{\mu}_k)\big)\Big)\nonumber\\
\!\!\!\!&=&\!\!\!\! \frac{1}{2^{m-1}}\sum_{\mu}
\delta\big(a_{k+1}^{ }\!\!-\!\!(a_{k+1}^0(\vec
x)\!\!+\!\!b^{\mu}_{k+1})\big) ...\delta\big(a_{m}^{
}\!\!-\!\!(a_{m}^0(\vec x)\!\!+\!\!\,b^{\mu}_{m})\big)\label{wy}
\end{eqnarray}\end{small}
where we use the fact that
 \begin{equation}
\sum_{a^{ }_1,...,a_k}\delta\big(a^{ }_1-(a_1^0(\vec
x)+b_1^{\mu})\big)...\delta\big(a_k-(a_k^0(\vec x)+b_k^{\mu})\big)
= 1
\end{equation}
whatever the value of $\vec{a}\,^0 (\vec x)+\vec{b}^{\mu}$ and for
all $\mu$, and $\vec x$. Now note that for any given $m-k$ element
bit string, $(a_{k+1},...,a_m)$, and $m-k$ element bit string,
$(a^0_{k+1}(\vec x),...,a^0_m(\vec x)\,)$ (for given $\vec x$),
there are  $2^{k-1}$ vectors $\vec b ^{\mu}$ such that
$(a_{k+1},...,a_m)=(a^0_{k+1}(\vec x),...,a^0_m(\vec
x))+(b^{\mu}_{k+1},...,b^{\mu}_{m})$. Thus, upon summing over
$\mu$, one finds, from eq. (\ref{wy}), that $P(a_{k+1},\ldots,a_m
| \vec x) = 2^{k-m}$  for all $a_{k+1},\ldots,a_m$ and for all
$\vec x$. The result is true for any $1\leq k\leq m$.  Thus no
non-trivial subgroup of the parties can signal to any other. The
correlations in eq. (\ref{Pmu}) are therefore non-signalling.
$\Box$.

\section{Conclusion}

Svetlichny \cite{Svet} and then others \cite{Seev,Col}
demonstrated that classical models which allow superluminal
communication within subsets of parties cannot reproduce all
multipartite quantum correlations. We have extended this approach
and showed that much more general classical communication patterns
than those considered in \cite{Seev,Col} cannot reproduce all
multipartite quantum correlations. We have shown how to describe
such communication patterns in terms of directed graphs. (For
instance the correlations considered in \cite{Svet,Seev,Col} are
described by separable graphs). Our main result is to prove that
correlations described by Partially Paired (PP) Graphs (see
definition 1) cannot reproduce all quantum correlations. PP graphs
are much more general than separable graphs, and therefore our
result shows that, in the multipartite setting, quantum
correlations are much more non-local than previously thought. To
obtain this result we carried out a detailed analysis of the
properties of the multiparty generalisation of the Svetlichny
inequality  for which the bounds attained by {\em lhv} models,
quantum mechanics, and completely non-local models were previously
known. We showed that the  correlations associated with PP graphs
attain the same bound as the {\em lhv} models, and therefore
cannot reproduce all quantum correlations. While for purposes of
exposition, we described the four party case in section \ref{V},
it should be noted that there are three party non-separable PP
correlations; for example $a_1(x_1,x_2,\wedge),
a_2(x_1,x_2,\wedge), a_3(\wedge,x_2,x_3)$.  In other words, the
phenomenon we have described in this paper, namely that quantum
mechanics is stronger than some in-separable correlations, appears
even for three party states. We have also found that another class
of correlations which are convex combinations of some of the
extreme points associated with Totally Paired (TP) graphs (see
definition 3) can both maximally violate the Svetlichny inequality
and be no-signalling. However this does not necessarily mean that
any TP graph has associated extreme points which can reproduce all
multipartite quantum correlations. Indeed the above results give
an essentially complete characterisation of how much different
classical communication patterns violate the Svetlichny
inequality. But there are many other Bell inequalities which can
be used to probe the non-locality of quantum correlations. It may
be that - using another Bell inequality as test - one can show
that some TP graphs represent correlations that cannot reproduce
all quantum correlations. On the other hand it may be that the
extreme points associated with any TP graph can reproduce all
quantum correlations. We leave this as an open question for future
research. Indeed the present work shows that the non-locality
present in multipartite quantum correlations is stronger, and
structurally richer, than previously thought.

\section*{Acknowledgements} We are grateful for support from the
EPSRC,  from project RESQ IST-2001-37559 of the IST-FET program of
the EC, from the Communaut\'e Fran{\c {c}}aise de Belgique under
grant ARC00/05-251 and from the IUAP program of the Belgian
government under grant V-18.


\begin{thebibliography}{30}
\bibitem{note that} Effectively there are two hidden variables. In
addition to $\lambda$, the variable $i\in\{1,2,3\}$ defines which
pair in the triple can communicate
\end{thebibliography}
\end{document}